\documentclass[preprint]{JHEP3} 

\usepackage{graphicx}
\usepackage{amsmath}
\usepackage{comment}

\newcommand{\cA}{{\cal A}}
\newcommand{\cD}{{\cal D}}
\newcommand{\cF}{{\cal F}}
\newcommand{\cN}{{\cal N}}
\newcommand{\cQ}{{\cal Q}}
\newcommand{\cU}{{\cal U}}

\newcommand{\rR}{{\rm R}}
\newcommand{\rRb}{\bar{{\rm R}}}

\newcommand{\hatbnu}{\widehat{\boldsymbol {\nu}}}
\newcommand{\vn}{ {\bf n} }

\newcommand{\cAb}{{\overline{\cal A}}}
\newcommand{\cFb}{{\overline{\cal F}}}
\newcommand{\cDb}{{\overline{\cal D}}}
\newcommand{\cUb}{{\overline{\cal U}}} 
 
\newcommand{\Tr}{\;{\rm Tr~}}
\newcommand{\hf}{\frac{1}{2}}

\def\phib{{\bar{\phi}}}

\def\etab{{\bar{\eta}}}
\def\psib{{\bar{\psi}}}
\def\chib{{\bar{\chi}}}

\def\nn{\nonumber}

\def\bec{\begin{center}}
\def\eec{\end{center}}
\def\beq{\begin{equation}}
\def\eeq{\end{equation}}
\def\bea{\begin{eqnarray}}
\def\eea{\end{eqnarray}}

\preprint{DESY-14-030}

\title{Two-dimensional ${\cal N} = (2, 2)$ Lattice Gauge Theories with Matter in Higher Representations}

\author{Anosh Joseph \\
John von Neumann Institute for Computing NIC, Platanenallee 6, 15738 Zeuthen, GERMANY\\
Deutsches Elektronen-Synchrotron DESY, Platanenallee 6, 15738 Zeuthen, GERMANY\\ 
E-mail: \email{anosh.joseph@desy.de}}

\abstract{We construct two-dimensional ${\cal N} = (2, 2)$ supersymmetric gauge theories on a Euclidean spacetime lattice with matter in the two-index symmetric and anti-symmetric representations of SU($N_c$) color group. These lattice theories preserve a subset of the supercharges exact at finite lattice spacing. The method of topological twisting is used to construct such theories in the continuum and then the geometric discretization scheme is used to formulate them on the lattice. The lattice theories obtained this way are gauge-invariant, free from fermion doubling problem and exact supersymmetric at finite lattice spacing. We hope that these lattice constructions further motivate the nonperturbative explorations of models inspired by technicolor, orbifolding and orientifolding in string theories and the Corrigan-Ramond limit.}


\begin{document}

\section{Introduction}
\label{sec:intro}

Supersymmetric Yang-Mills (SYM) theories are interesting classes of theories by themselves. They also serve as starting points for constructions of many phenomenologically relevant models. These theories can come with nonperturbative sectors that are less tractable analytically. Having a nonperturbative formulation of supersymmetric gauge theories would certainly advance our goal toward understanding their rich structure. Supersymmetric gauge theories constructed on a Euclidean spacetime lattice would provide a first principles approach to study the nonperturbative regimes of these theories. For certain classes of SYM theories with extended supersymmetries there are two approaches readily available to us for the constructions of their supersymmetric cousins. They are called the methods of topological twisting and orbifolding \cite{Kaplan:2002wv, Cohen:2003xe, Cohen:2003qw, Catterall:2004np, Catterall:2005fd, Kaplan:2005ta, Unsal:2006qp, Catterall:2007kn, Catterall:2009it, Catterall:2011pd, Joseph:2011xy, Catterall:2013roa}. They both give rise to identical lattices\footnote{There also exist other complementary approaches to the problem of lattice supersymmetry \cite{Sugino:2003yb,Sugino:2004qd, D'Adda:2005zk, D'Adda:2007ax, Kanamori:2008bk, Hanada:2009hq, Hanada:2010kt, Hanada:2010gs, Hanada:2011qx, Matsuura:2014pua}.}. 

Supersymmetric lattices have been constructed for several classes of SYM theories \cite{Cohen:2003xe, Cohen:2003qw, Catterall:2004np, Catterall:2005fd, Kaplan:2005ta} including the well known $\cN=4$ SYM theory\footnote{Several aspects of this theory have been explored numerically in the recent past \cite{Catterall:2014vka, Weir:2013zua, Catterall:2012yq, Catterall:2011aa, Mehta:2011ud, Galvez:2012sv, Catterall:2010ya, Catterall:2010fx}}. There have been a few extensions of these formulations by incorporating matter fields in the adjoint and fundamental representations of the color group \cite{Endres:2006ic, Giedt:2006dd, Giedt:2011zza, Matsuura:2008cfa, Joseph:2013jya}. Some of them have also been extended to incorporate product gauge groups, resulting in supersymmetric quiver gauge theories on the lattice \cite{Matsuura:2008cfa, Joseph:2013jya, Joseph:2013bra}. 

In this paper, we construct two-dimensional $\cN = (2, 2)$ supersymmetric lattice gauge theories with matter fields transforming in the higher representations of SU($N_c$) color group. These theories are constructed using the following procedure. We begin with a two-dimensional Euclidean SYM theory possessing eight supercharges. Such a theory can be obtained from dimensionally reducing the four-dimensional Euclidean $\cN = 2$ SYM. The fields and supersymmetries of this theory are then topologically twisted to obtain a continuum theory, which is compatible with lattice discretization. The next step is to extend this theory such that it becomes a supersymmetric quiver gauge theory with two nodes and with gauge group SU($N_c$) $\times$ SU($N_f$). This can be achieved by replicating the continuum twisted theory and then changing the group representation of an appropriate subset of the field content of the theory from adjoint to the product representations $(N_c, \overline{N}_f)$ and $(\overline{N}_c, N_f)$, with $N_c$ and $N_f$ being the fundamental representations of SU($N_c$) and SU($N_f$) respectively. We note that such a construction could be obtained from an appropriate orbifold projection of an SU($N_c + N_f$) parent theory but we will not take that path of construction in this paper. We are interested in constructing $\cN = (2, 2)$ lattice theories with two-index representations for the matter fields. From the point of view of the orbifold constructions one has to find an appropriate projection method to get the quiver theories with desired two-index matter representations. However, in this paper we will not seek such a projection method but instead impose the projection {\it by hand} to construct the continuum theories with two-index matter and then proceed to construct lattice versions of these theories that preserve a subset of the supercharges exact on the lattice. Denoting the desired two-index representation by $\rR$ and the corresponding complex conjugate representation by $\rRb$, our starting point would be a quiver theory with matter in the representations $(\rR, \rRb)$ and $(\rRb, \rR)$. The adjoint fields of the quiver theory live on the nodes while the fields in the product representations live on the links connecting the nodes of the quiver. To construct two-dimensional $\cN = (2, 2)$ lattice gauge theories with two-index matter, we freeze the theory on one of the nodes of the quiver and also an appropriate set of matter fields linking the two nodes. After this restriction we have an SU($N_c$) gauge theory containing matter fields in the two-index representation and with SU($N_f$) flavor symmetry. Note that such a restriction of the fields is not in conflict with supersymmetry. Such constructions have been carried out by Matsuura \cite{Matsuura:2008cfa} and also in Ref. \cite{Joseph:2013jya} to formulate lattice gauge theories with matter fields in the fundamental representation. The continuum theories constructed this way, with two-index matter, can then be placed on the lattice using the method of geometric discretization. The resultant lattice theories contain adjoint fields living on the p-cells of a two-dimensional square lattice and two-index matter fields living on the sites of the same lattice. The lattice theories constructed this way enjoy gauge-invariance, freedom from fermion doublers and exact supersymmetry at finite lattice spacing.   

This paper is organized as follows. In Sec. \ref{sec:sym_2d_adj_matter} we write down the topologically twisted action and scalar supersymmetry transformations of the two-dimensional eight supercharge SYM with gauge group SU($N_c$). In Sec. \ref{sec:sym_2d_2index_matter} we construct the two-dimensional $\cN = (2, 2)$ gauge theory with matter fields transforming in the two-index (symmetric and anti-symmetric) representations of SU($N_c$). Our construction gives rise to matter fields with SU($N_f$) flavor symmetry. We construct the lattice theory with two-index matter in Sec. \ref{sec:lattice_theories} using the method of geometric discretization. In Sec. \ref{sec:fine_tuning} we briefly discuss the fine tuning and simulation details of the lattice theories. We end with discussion, comments and motivations to study such lattice models in Sec. \ref{sec:discussion_comments}.

\section{$\cN = (2, 2)$ Theories with Adjoint Matter}
\label{sec:sym_2d_adj_matter}

We begin with writing down the topologically twisted action of the two-dimensional Euclidean SYM theory with eight supercharges. The twisted action of the theory is obtained by decomposing the fields and supercharges of the original theory under a new rotation group called the twisted rotation group \cite{Witten:1988ze}. In our case, the twisted rotation group is the diagonal subgroup of the product of the two-dimensional Euclidean Lorentz rotation group SO($2$)$_{\rm E}$ and the SO($2$) subgroup of the R-symmetry group of the original theory. The action of the two-dimensional theory is
\beq
\label{eq:2d_Q8_adj}
S = S_{{\rm SYM}}^{\cN=(2, 2)} + S_{{\rm adj~matter}},
\eeq
where the first piece is the action of the two-dimensional $\cN=(2, 2)$ SYM
\beq
\label{eq:2d_Q4_action}
S_{{\rm SYM}}^{\cN=(2, 2)} = \frac{1}{g^2} \int d^2x \Tr \Big(-\cFb_{mn} \cF_{mn} + \frac{1}{2}[\cDb_m, \cD_m]^2 - \chi_{mn} \cD_m \psi_n - \eta \cDb_m \psi_m \Big),
\eeq
and the second piece contains the matter part in the adjoint representation
\bea
S_{{\rm adj~matter}} &=& \frac{1}{g^2} \int d^2 x~\Tr \Big(-2(\cDb_m\phib)(\cD_m\phi) + [\cDb_m, \cD_m][\phib, \phi] + \etab \cD_m \psib_m \nn \\
&&+ \chib_{mn}\cDb_m \psib_n - \eta [\phib, \etab] - \psi_m [\phi, \psib_m] - \hf \chi_{mn} [\phib, \chib_{mn}] + \hf [\phib, \phi]^2 \Big).
\eea

Here $m, n = 1, 2$ and $g$ is the coupling parameter of the theory. All fields are transforming in the adjoint representation of the SU($N_c$) gauge group\footnote{For the adjoint representation we use the anti-hermitian basis formed by the SU($N_c$) generators in the defining representation; the $N_c \times N_c$ matrices $T^a$ with $a = 1, 2, \cdots, N_c^2 - 1$. They have the normalization $\Tr (T^a T^b) = -\hf \delta^{ab}$. In this paper we express all group theoretical weights in terms of the defining representation.}.

After topological twisting, the fermions and supercharges transform as integer spin representations of the twisted rotation group. The fermionic degrees of freedom of the twisted theory are p-forms with p$=0, 1, 2$. They are labeled as $\{\eta$, $\psi_m$, $\chi_{mn}$, $\etab$, $\psib_m$, $\chib_{mn}\}$. Similarly, the twisted supercharges can also be packaged as a set of p-forms. The untwisted theory contains four scalars. After twisting, two of the scalars combine to form a two-dimensional vector $B_m$ under the twisted rotation group. Since there are two vector fields in the twisted theory, $A_m$ and $B_m$, and they both transform the same way under the twisted rotation group, it is natural to combine them to form a complexified gauge field, which we label as $\cA_m$, and write down the twisted theory in a compact way. Thus the twisted theory contains two complexified gauge fields
\bea
\cA_m \equiv A_m + iB_m,~~~\cAb_m \equiv A_m - iB_m.
\eea

Such a construction leads to complexified covariant derivatives in the theory. They are defined by
\bea
\label{eq:cov_der_adj1}
\cD_m ~ &=& \partial_m ~ + ~[\cA_m, ~~~ ] = \partial_m ~ + ~[A_m + iB_m, ~~ ], \\
\label{eq:cov_der_adj2}
\cDb_m ~ &=& \partial_m ~ + ~[\cAb_m, ~~~ ] = \partial_m ~ + ~[A_m - i B_m, ~~ ].
\eea
The complexification of gauge fields also results in complexified field strengths 
\beq
\cF_{mn} = [\cD_m, \cD_n]\quad {\rm and} \quad \cFb_{mn} = [\cDb_m, \cDb_n].
\eeq
The twisted theory also contains two scalars $\phi$ and $\phib$.

Among the twisted supercharges, the scalar (0-form) supercharge $\cQ$ is important for us. This supercharge is nilpotent, $\cQ^2 = 0$. It does not produce any infinitesimal translations and thus we can transport this subalgebra of the twisted supersymmetry algebra to the lattice. In the twisted supersymmetry algebra one also finds that the momentum is the $\cQ$-variation of something, which makes plausible the statement that one can write the energy-momentum tensor, and the entire action in a $\cQ$-exact form. Thus a lattice action constructed in a $\cQ$-exact form is trivially invariant under the scalar supercharge. In summary, we can use the process of twisting to construct a lattice action that respects at least one supersymmetry exact on the lattice. The lattice theories formulated using twisted fermions are free from the fermion doubling problem, owing to the property that the twisted fermions are geometric in nature (p-forms) and thus they can be mapped one-to-one on to the lattice from continuum \cite{Rabin:1981qj, Becher:1982ud, Banks:1982iq, Aratyn:1984bd}.

The scalar supercharge generates the following nilpotent supersymmetry transformations on the twisted fields
\begin{align}
\label{eq:susy_adj_2d_Q8}
\cQ \cA_m& =\psi_m,&\cQ \cAb_m& =0,\\
\cQ \phi& =\etab,&\cQ \phib& =0,\\
\cQ \eta& = d,&\cQ \etab& =0,\\
\cQ \psi_m& =0,&\cQ \psib_m& = \cDb_m \phib,\\
\cQ \chi_{mn}& =-[\cDb_m, \cDb_n],&\cQ \chib_{mn} &=0, \\
\cQ d &= 0,
\end{align}
where $d$ is an auxiliary field introduced for the off-shell completion of the algebra. It obeys the constraint $d = \sum_m [\cDb_m, \cD_m] + [\phib, \phi]$. Note that while writing down the $\cQ$-exact action in Eq. (\ref{eq:2d_Q8_adj}) the auxiliary field has been integrated out to express the action in the on-shell form.
 
\section{$\cN = (2, 2)$ Theories with Two-index Matter}
\label{sec:sym_2d_2index_matter}

We are interested in constructing two-dimensional $\cN = (2, 2)$ lattice gauge theories coupled to matter fields transforming in the two-index representations of SU($N_c$) gauge group. To construct such theories we follow the procedure similar to the one given in Ref. \cite{Matsuura:2008cfa}. We begin with the action of the theory given in Eq. (\ref{eq:2d_Q8_adj}). The next step is to make two copies of the theory to construct a quiver gauge theory containing two nodes and with gauge group SU($N_c$) $\times$ SU($N_f$). The nodes are labeled by $N_c$ and $N_f$. Each node contains a copy of the two-dimensional $\cN = (2, 2)$ SYM, with gauge group SU($N_c$) for the $N_c$-node and SU($N_f$) for the $N_f$-node. This type of construction, through orbifold projection, usually leads to matter in bi-fundamental representations $(\overline{N}_c, N_f)$ and $(N_c, \overline{N}_f)$ of SU($N_c$) $\times$ SU($N_f$). However, in this paper we consider matter fields in the two-index representation of SU($N$) transforming as $(\rR, \rRb)$ and $(\rRb, \rR)$ under the product gauge group, with $\rR$ being the desired two-index (symmetric or anti-symmetric) representation and $\rRb$ the corresponding complex conjugate representation\footnote{We note that in string theory one can obtain the two-index representation of SU($N$) by performing the orientifold projection to adjoint of SO($2N$) or Sp($2N$). The theory we consider in this paper would be obtained in the same manner. I thank the referee for pointing this out.}. One could, in principle, look for a suitable projection that leads to matter in two-index representations but here we impose such a requirement {\it by hand}. The matter fields of the quiver theory now live on the links connecting the two nodes of the quiver. The next step is to freeze one of the nodes of the theory, say the $N_f$-node and make one set of the link fields non-dynamical by hand. The resulting theory is a two-dimensional $\cN = (2, 2)$ gauge theory with matter in the two-index representation of SU($N_c$) gauge group and with SU($N_f$) flavor symmetry. The theory with two-index matter still contains the scalar supercharge and it can be discretized on a lattice. In the lattice theory, the fields of the SYM multiplet are placed on the p-cells of the lattice. The matter fields of the lattice theory occupy the sites of the lattice. Such a prescription leads to a gauge-invariant lattice theory preserving a subset of supercharges exact on the lattice.

The action of the continuum theory with two-index matter has the following form
\beq
S = S_{{\rm SYM}}^{\cN=(2, 2)} + S_{{\rm 2I~matter}},
\eeq
where the first part is the same as the one given in Eq. (\ref{eq:2d_Q4_action}). It contains the set of fields $\{\cA_m$, $\cAb_m$, $\eta$, $\psi_m$, $\chi_{mn}\}$ transforming in the adjoint representation of SU($N_c$). The matter part of the action contains fields transforming in the adjoint, two-index and two-index complex conjugate representations of SU($N_c$) and is given by 
\bea
S_{{\rm 2I~matter}} &=& \frac{1}{g^2} \int d^2 x~\Tr \Big(2 \phib^\alpha (\cDb_m\cD_m) \phi^\alpha - [\cDb_m, \cD_m] (\phi^\alpha \phib^\alpha) + \psib^\alpha_m \cD_m \etab^\alpha \nn \\
&&+ \chib^\alpha_{mn}\cDb_m \psib^\alpha_n + \eta \etab^\alpha \phib^\alpha - \psi_m \phi^\alpha \psib^\alpha_m + \hf \chi_{mn} \chib^\alpha_{mn} \phib^\alpha \nn \\
&&+ \hf (\phi^\alpha \phib^\alpha)^2 + \hf (\phib^\alpha \phi^\alpha)^2 \Big),~~~~
\eea
with $\alpha$ an index labeling the SU($N_f$) flavor symmetry.

The set of fields $\{\phi^\alpha, \etab^\alpha, \chib^\alpha_{12}\}$ transform in the two-index representation $\rR$ and the set of fields $\{\phib^\alpha, \psib^\alpha_1, \psib^\alpha_2\}$ transform in the complex conjugate representation $\rRb$. 

For a generic field $\Psi$ in the two-index representation, the action of the covariant derivative is given by
\bea
\cD_m \Psi^{\rR} &=& \partial_m \Psi^{\rR} + \cA_m^{\rR} \Psi^{\rR},
\eea
while for a field $\overline{\Psi}$ in the complex conjugate representation
\bea
\cD_m \overline{\Psi}^{\rRb} &=& \partial_m \overline{\Psi}^{\rRb} + \overline{\Psi}^{\rRb} \cA_m^{\rRb},
\eea
with $\cA_m^{\rR} = \cA^a_m (t^a)^{\rR}$, $\cA_m^{\rRb} = \cA^a_m (t^a)^{\rRb}$ and $a = 1, 2, \cdots, N_c^2 - 1$. 

The generators of the two-index representation $(t^a)^{\rR}$ and $(t^a)^{\rRb}$ can be constructed from the generators of the defining representation of SU($N_c$) using appropriate projection operators that pick out the anti-symmetric or symmetric part of the second-rank tensor representation of SU($N$) \cite{Cvitanovic:1976am}. Denoting $(t^a)^{\rm 2A}$ and $(t^a)^{\rm 2S}$ as the generators for the two-index anti-symmetric (2A) and symmetric (2S) representations, respectively, we can write down their expressions in terms of the generators of the fundamental representation of SU($N_c$). 

Using the projection operators $P^A = \hf (\delta_{ik}\delta_{jl} - \delta_{jk}\delta_{il})$, $i < j,~k < l$, and $P^S = \hf (\delta_{ik}\delta_{jl} + \delta_{jk}\delta_{il})$, $i \leq j,~k \leq l$, where $i, j, k, l = 1, 2, \cdots N_c$, we pick out the two-index anti-symmetric and symmetric parts of the SU($N_c$) representation. The generator $(t^a)^{\rm 2A}$ is defined as
\bea
(t^a)^{\rm 2A} &=& (t^a)^{\rm 2A}_{ij, kl} = \hf [(T^a)_{ik}\delta_{jl} - (T^a)_{jk}\delta_{il} + \delta_{ik}(T^a)_{jl} - \delta_{jk}(T^a)_{il}],
\eea
and the generator $(t^a)^{\rm 2S}$
\bea
(t^a)^{\rm 2S} &=& (t^a)^{\rm 2S}_{ij, kl} = \hf [(T^a)_{ik}\delta_{jl} + (T^a)_{jk}\delta_{il} + \delta_{ik}(T^a)_{jl} + \delta_{jk}(T^a)_{il}].
\eea
Indeed they have the correct dimensions and indices of the respective representations: $d_{2A/2S} = \hf N_c(N_c \mp 1)$ and $T^{2A/2S} = \hf (N_c \mp 2)$. The generators of the complex conjugate representations are obtained from $(t^a)^{\rRb} = \left( (t^a)^{\rR} \right)^*$.

\section{Lattice Theories}
\label{sec:lattice_theories}

The $\cN = (2, 2)$ gauge theories with matter constructed above can be discretized on a Euclidean spacetime lattice using the geometric discretization scheme formulated in Refs. \cite{Catterall:2007kn, Damgaard:2007be, Damgaard:2008pa}. The complexified gauge fields of the continuum theory, $\cA_m(x)$ are mapped to appropriate complexified Wilson links $\cU_m(\vn)$ defined at a location on the two-dimensional square lattice denoted by the integer vector $\vn$. These link fields are associated with unit length vectors in the coordinate directions $\hatbnu_m$ from the site $\vn$. The components of the fermion field $\psi_m(\vn)$ live on the same oriented links as that of their bosonic superpartners $\cU_m(\vn)$. The field $\eta(\vn)$ is placed on the site $\vn$. The components of the field $\chi_{mn}(\vn)$ are placed on a set of diagonal face links with orientation $\vn + \hatbnu_m + \hatbnu_n \to \vn$.

We can write down the gauge transformation rules for the lattice fields in the adjoint representation respecting the p-cell and orientation assignments on the lattice. For $G(\vn) \in$ SU($N_c$), we have the following gauge transformation prescriptions \cite{Aratyn:1984bd, Damgaard:2008pa}

\bea
\label{eq:gauge_latt_1}
\cU_m(\vn) &\rightarrow& G(\vn) \cU_m(\vn) G^{\dagger}(\vn + \hatbnu_m), \\ 
\cUb_m(\vn) &\rightarrow& G(\vn + \hatbnu_m) \cUb_m(\vn)G^{\dagger}(\vn), \\ 
\eta(\vn) &\rightarrow& G(\vn) \eta(\vn)G^{\dagger}(\vn), \\ 
\psi_m(\vn) &\rightarrow& G(\vn) \psi_m(\vn)G^{\dagger}(\vn + \hatbnu_m), \\ 
\chi_{mn}(\vn) &\rightarrow& G(\vn + \hatbnu_m + \hatbnu_n) \chi_{mn}(\vn)G^{\dagger}(\vn).
\label{eq:gauge_latt_2}
\eea

The covariant derivatives in the continuum are mapped to covariant difference operators on the lattice. The covariant derivatives $\cD_m$ ($\cDb_m$) of the continuum theory become forward and backward covariant difference operators $\cD^{(+)}_m$ ($\cDb^{(+)}_m$) and $\cD^{(-)}_m$ ($\cDb^{(-)}_m$), respectively of the lattice theory. The forward and backward covariant difference operators act on the adjoint lattice fields in the following way
\bea
\label{eq:cov_diff_op_adj_1}
\cD_m^{(-)} \psi_m (\vn) &=& \cU_m(\vn) \psi_m(\vn) - \psi_m(\vn - \hatbnu_m) \cU_m(\vn - \hatbnu_m), \\
\cD_m^{(+)} \psi_n (\vn) &=& \cU_m(\vn) \psi_n(\vn + \hatbnu_m) - \psi_n(\vn) \cU_m(\vn + \hatbnu_n), \\
\cDb_m^{(-)} \psi_m (\vn) &=& \psi_m(\vn) \cUb_m(\vn) - \cUb_m(\vn - \hatbnu_m) \psi_m(\vn - \hatbnu_m), \\
\label{eq:cov_diff_op_adj_2}
\cDb_m^{(+)} \chi_{nq}(\vn) &=& \chi_{nq}(\vn + \hatbnu_m) \cUb_m(\vn) - \cUb_m(\vn + \hatbnu_n + \hatbnu_q) \chi_{nq}(\vn).
\eea

These expressions reduce to the corresponding continuum results for the adjoint covariant derivative in the naive continuum limit. They also transform under gauge transformations like the corresponding lattice link field carrying the same indices. Such a discretization prescription leads to a lattice action with terms corresponding to gauge-invariant closed loops. 

The field strength on the lattice is given by the expression $\cF_{mn}(\vn) = \cD^{(+)}_m \cU_n(\vn)$. It is automatically antisymmetric in its indices and also it transforms like a 2-form on the lattice. It has an orientation opposite to that of the field $\chi_{mn}$ on the unit cell.

We also need to define the action of the covariant difference operators on the lattice fields transforming in the two-index representations. The two-index matter fields of the lattice theory live on the  sites of the lattice. They transform in the two-index representations of SU($N_c$). We write down the following set of rules for the action of the covariant derivatives on fields in the two-index representations. 

The covariant difference operator acts on the lattice variables in the two-index representation $\rR$ the following way:
\bea
\cD_m^{(+)}\Phi^{\rR}(\vn) &\equiv& \cU_m(\vn)\Phi^{\rR}(\vn + \hatbnu_m) - \Phi^{\rR}(\vn), \\
\cDb_m^{(+)}\Phi^{\rR}(\vn) &\equiv& \Phi^{\rR}(\vn + \hatbnu_m) - \cUb_m(\vn)\Phi^{\rR}(\vn), \\
\cD_m^{(-)}\Phi^{\rR}(\vn) &\equiv& \cD^{(+)}_m \Phi^{\rR}(\vn - \hatbnu_m),~~\cDb_m^{(-)}\Phi^{\rR}(\vn) \equiv \cDb^{(+)}_m \Phi^{\rR}(\vn - \hatbnu_m).
\eea

For lattice variables in the complex conjugate representation $\rRb$ we have the following set of rules for the action of the covariant difference operator:
\bea
\cD_m^{(+)}\Phi^{\rRb}(\vn) &\equiv& \Phi^{\rRb}(\vn + \hatbnu_m) - \Phi^{\rRb}(\vn)\cU_m(\vn), \\
\cDb_m^{(+)}\Phi^{\rRb}(\vn) &\equiv& \Phi^{\rRb}(\vn + \hatbnu_m)\cUb_m(\vn) - \Phi^{\rRb}(\vn), \\
\cD_m^{(-)}\Phi^{\rRb}(\vn) &\equiv& \cD^{(+)}_m \Phi^{\rRb}(\vn - \hatbnu_m),~~\cDb_m^{(-)}\Phi^{\rRb}(\vn) \equiv \cDb^{(+)}_m \Phi^{\rRb}(\vn - \hatbnu_m).
\eea

We see that the above expressions reduce to the corresponding continuum covariant derivatives in the naive continuum limit by letting $\cU_m(\vn) = {\mathbf I} + \cA_m(\vn) + \cdots$ and $\cUb_m(\vn) = {\mathbf I} - \cAb_m(\vn) + \cdots$.

The fields in the two-index and its complex conjugate representations are mapped on to lattice sites, with the gauge transformations
\bea
\Phi^{\rR}(\vn) &\rightarrow& G(\vn) \Phi^{\rR}(\vn), \\
\Phi^{\rRb}(\vn) &\rightarrow& \Phi^{\rRb}(\vn) G^\dagger(\vn).
\eea

\begin{figure}
\begin{center}
\includegraphics[width=0.5\textwidth]{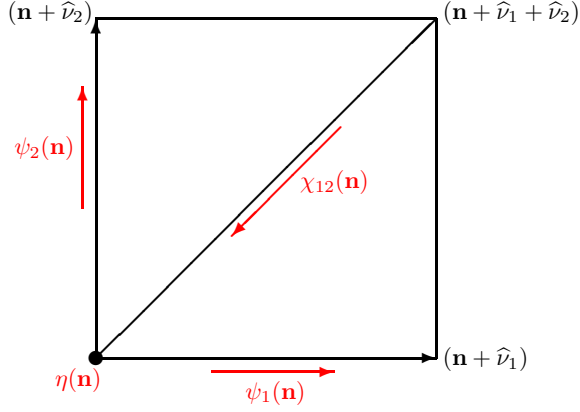}
\end{center}
\caption{\label{fig:2dlatt}The unit cell of the two-dimensional $\cN = (2, 2)$ lattice SYM with orientation assignments for the twisted fermions. The complexified bosons $\cU_m(\vn)$ follow the same orientations and link assignments as that of their superpartners $\psi_m(\vn)$. The lattice field strength $\cF_{mn}(\vn)$ is placed on the diagonal link with the orientation opposite to that of the field $\chi_{mn}(\vn)$.}
\end{figure}

We also note that the method of geometric discretization maps the continuum fields on to the lattice one-to-one and thus the lattice theories constructed this way are free from the fermion doubling problem \cite{Rabin:1981qj, Becher:1982ud, Banks:1982iq, Aratyn:1984bd}. The placement and orientations of the twisted fields on the lattice respect the scalar supersymmetry and gauge symmetry of the lattice theory. The unit cell of the two-dimensional lattice theory is given in Fig. \ref{fig:2dlatt}.

The scalar supersymmetry acts on the lattice variables the following way
\begin{align}
\cQ \cU_m(\vn)& =\psi_m(\vn),&
\cQ \cUb_m(\vn)& =0,\\
\cQ \phi^\alpha(\vn)& =\etab^\alpha(\vn),&
\cQ \phib^\alpha(\vn)& =0,\\
\cQ \eta(\vn)& = d(\vn),&\cQ \etab^\alpha(\vn)& =0,\\
\cQ \psi_m(\vn)& =0,&\cQ \psib^\alpha_m(\vn)& = \Big(\cDb^{(+)}_m\phib^\alpha\Big)(\vn),\\
\cQ \chi_{mn}& =-\Big(\overline{\cD^{(+)}_m\cU_n}\Big)(\vn),&\cQ \chib^\alpha_{mn}(\vn) &=0, \\
\cQ d(\vn) &= 0.
\end{align}

Once we have the rules to map the continuum twisted theory to the lattice it is easy to write down the lattice action. The two-dimensional $\cN = (2, 2)$ gauge theory with two-index matter has the following form on the lattice
\beq
S = S_{{\rm SYM}}^{\cN=(2, 2)} + S_{\rm 2I~matter},
\eeq
with
\bea
S_{{\rm SYM}}^{\cN=(2, 2)} &=& \frac{1}{g^2} \sum_{\vn} \Tr \Big\{\Big(\cUb_n(\vn + \hatbnu_m)\cUb_m(\vn) - \cUb_m(\vn + \hatbnu_n)\cUb_n(\vn)\Big)\nn \\
&&\times \Big(\cU_m(\vn)\cU_n(\vn + \hatbnu_m) - \cU_n(\vn)\cU_m(\vn + \hatbnu_n)\Big) \nn \\
&&+ \hf \Big(\cU_m(\vn)\cUb_m(\vn) - \cUb_m(\vn - \hatbnu_m)\cU_m(\vn - \hatbnu_m)\Big)^2 \nn \\
&&+\hf(\delta_{mq}\delta_{nr} - \delta_{mr}\delta_{nq}) \chi_{mn}(\vn) \Big(\cU_q(\vn)\psi_r(\vn + \hatbnu_q) - \psi_r(\vn)\cU_q(\vn + \hatbnu_r)\Big) \nn \\
&&+ \eta(\vn)\Big(\psi_m(\vn)\cUb_m(\vn) - \cUb_m(\vn - \hatbnu_m)\psi_m(\vn - \hatbnu_m)\Big)\Big\},
\eea
and
\bea
S_{{\rm 2I~matter}} &=& \frac{1}{g^2} \sum_{\vn} \Tr \Big(2 \phib^\alpha(\vn) \cDb^{(-)}_m \cD^{(+)}_m \phi^\alpha(\vn) - \Big(\cDb^{(-)}_m \cU_m(\vn)\Big) \Big(\phi^\alpha(\vn) \phib^\alpha(\vn)\Big) \nn \\
&&+ \psib^\alpha_m(\vn) \cD^{(+)}_m \etab^\alpha(\vn) + \chib^\alpha_{mn}(\vn) \cDb^{(+)}_m \psib^\alpha_n(\vn) + \eta(\vn) \etab^\alpha(\vn) \phib^\alpha(\vn) \nn \\
&&- \psi_m(\vn) \phi^\alpha(\vn + \hatbnu_m) \psib^\alpha_m(\vn) + \hf \chi_{mn}(\vn) \chib^\alpha_{mn}(\vn) \phib^\alpha(\vn + \hatbnu_m + \hatbnu_n) \nn \\
&&+ \hf (\phi^\alpha(\vn) \phib^\alpha(\vn))^2 + \hf (\phib^\alpha(\vn) \phi^\alpha(\vn))^2 \Big).
\eea
 
We note that the lattice action written above is $\cQ$-supersymmetric, gauge-invariant, local and free from the fermion doublers. 
 
\section{Fine Tuning and Simulation on the Lattice}
\label{sec:fine_tuning}

In this paper, we have constructed the lattice actions of two-dimensional $\cN = (2, 2)$ gauge theories with matter fields transforming in the higher representations of SU($N_c$) gauge group. On the lattice, radiative corrections could induce dangerous operators that could violate the Euclidean rotation symmetry and supersymmetry invariance of the theory as we approach the continuum limit. We would like to know whether the above constructed supersymmetric lattice theories are free from fine tuning as the continuum limit is approached. We could check this at least perturbatively by using a power counting analysis.

We note that the allowed operators on the lattice have to respect the $\cQ$-supersymmetry, R-symmetry, flavor symmetry, gauge symmetry and the point group symmetry $S_2$. For a given operator $O^{(p)}$ with mass dimension $p$, the counterterms in the lattice action can take the following general form:
\beq
\delta S = \frac{1}{g^2} \int d^2x~ C_p O^{(p)},
\eeq
where $g$ is the coupling parameter, which has mass dimension 1. The coefficient $C_p$ denotes the contributions from the loop expansion
\beq
C_p = a^{p - 4} \sum_l c_l (g^2 a^2)^l,
\eeq
with $l$ counting the number of loops in a perturbative expansion and $a$ denoting the lattice spacing. The dimensionless coefficient $c_l$ can depend at most logarithmically on the lattice spacing. Assigning the following mass dimensions to the fields $[\Phi] = 1$, $[\Psi] = \frac{3}{2}$ and $[\cQ] = \hf$, where $\Phi$ and $\Psi$ denote the twisted bosons and fermions respectively, we have the following generic form for the operator
\bea
O^{(p)} &=& \cQ \Tr \Big(f(\Phi) g(\Psi)\Big).
\eea
We see that such operators are annihilated by the $\cQ$ supersymmetry and thus they are $\cQ$-invariant. Gauge-invariance on the lattice requires that all the fields in the operator must be oriented such that the operator should correspond to the trace of a closed loop on the lattice.

The coefficient of any dangerous operator should vanish in the limit $a \rightarrow 0$ to have a nice continuum limit for the theory. Since radiative corrections start at one-loop, $l=1$, we need to check whether the lattice theories allow operators with mass dimension $p \leq 3$ that respect all the lattice symmetries. Naively we can write down the following set of such operators
\beq
\Big\{ \Phi, \Phi^2, \Phi^3, (\Phi \partial \Phi), (\Psi\Psi), \cQ (\Phi \Psi) \Big\}.
\eeq

Among the set of possible operators, we see that naively there could be scalar mass terms, $\Tr \Phi^2$, induced through radiative corrections. For a dimension 2 operator, we have
\beq
\delta S^{(p=2)} = \int d^2x~\Big( c_1 + c_2 g^2 a^2 + \cdots \Big) O^{(p=2)},
\eeq
telling us that scalar mass terms could be induced at one-loop, which can have at most a logarithmic divergence, and higher loop vanishing contributions. However it appears that scalar mass terms cannot respect all the symmetries of the lattice and we conclude that they cannot be generated via radiative corrections.

We see that dimension 3 operators, including the fermionic mass terms $\Tr(\Psi\Psi)$, could not be induced at any loop
\beq
\delta S^{(p=3)} = \int d^2x~\Big( c_1 a + c_2 g^2 a^3 + \cdots \Big) O^{(p=3)}.
\eeq
Thus fermion bilinear counterterms cannot be generated radiatively.

We note that the lattice theories constructed above exhibit flat directions (a general property of extended supersymmetric theories) and they give rise to instabilities while performing lattice simulations\footnote{See \cite{Catterall:2011cea} for the details of a numerical code to simulate the topologically twisted SYM theories.}. A way to control them in lattice simulations is to introduce suitable gauge invariant, but $\cQ$ non-invariant, potential terms (similar to the approach described in Ref. \cite{Catterall:2011aa}) {\it by hand}. Another point to note is that these theories might also suffer from fermion sign problem. One has to explore the existence of sign problem in these theories to boost confidence that these lattice constructions can be used successfully to explore non-perturbative aspects of two-dimensional $\cN = (2, 2)$ gauge theories with matter formulated here\footnote{In Ref. \cite{Catterall:2011aa} it has been shown through lattice simulations that the SYM theories with four and sixteen supercharges in two dimensions do not suffer from fermion sign problem.}.

\section{Discussion and Comments}
\label{sec:discussion_comments}

In this paper, we have detailed the lattice constructions of two-dimensional $\cN = (2, 2)$ supersymmetric gauge theories coupled with matter fields in higher representations of gauge group SU($N_c$). The process of topological twisting allows us to write down the theories in the continuum in a way compatible with lattice discretization. We used the method of geometric discretization to formulate these theories on a two-dimensional Euclidean spacetime lattice. The lattice theories constructed this way are local, gauge-invariant, doubler free and retain one supercharge exact on the lattice. The matter fields of these theories are in the two-index symmetric and antisymmetric representations of SU($N_c$) gauge group. The process of un-gauging one of the nodes of the quiver theory, while constructing the desired theory with two-index matter, also results in matter fields to have SU($N_f$) flavor symmetry. 

We also note that one could construct $\cN = (2, 2)$ supersymmetric quiver lattice gauge theories with matter fields transforming in the product representations $(\rR, \rRb)$ and $(\rRb, \rR)$ with $\rR$ the two-index representation (symmetric or anti-symmetric) and $\rRb$ the corresponding complex conjugate representation. One could also construct lattice theories with eight supercharges coupled to matter fields in two-index representations using the approach detailed in this paper. For such constructions the starting point would be the sixteen supercharge Yang-Mills theory in four-dimensions. We also note that it would be interesting to construct the theories detailed here through a suitable orbifold projection method.

When the gauge group is SU($3$), we see that the 2A representation is the same as the anti-fundamental representation. So, for $N_c=3$, this is also a construction for the SU($3$) theory with anti-fundamental fermions. It would be interesting to explore the lattice theories formulated in this paper in the context of the Corrigan-Ramond (CR) limit \cite{Corrigan:1979xf}, which is another unique way to take the large-$N_c$ limit of a theory. In the QCD theory originally proposed by Corrigan and Ramond, some quark flavors live in the two-index anti-symmetric representation and others still in the fundamental representation of SU($N_c$). When the theory has three colors, the CR limit is identical to QCD (up to charge conjugation). We note that for large-$N_c$ the CR limit and the 't Hooft limit are very different. Recently, this particular idea of the large-$N_c$ limit was revived by Armoni, Shifman and Veneziano \cite{Armoni:2003gp, Armoni:2003fb}. They discovered a remarkable relationship between certain sectors of the two-index anti-symmetric and symmetric theories at large-$N_c$ and sectors of SYM theories. We also hope that the lattice constructions detailed here would further motivate nonperturbative constructions and explorations in certain theories with technicolor \cite{Dietrich:2005jn}. The lattice theories formulated here would be interesting in the context of string theory. There one can obtain the two-index representation of SU($N_c$) by performing the orientifold projection to adjoint of SO($2N_c$) or Sp($2N_c$). The continuum (and thus lattice) theories we consider in this paper would be obtained in the same manner. Symmetric and antisymmetric representations of matter are interesting in the context of chiral supersymmetric gauge theories as these representations serve as building blocks of such theories. In particular, many models with dynamical supersymmetry breaking are based on anti-symmetric representations \cite{Affleck:1983vc, Meurice:1984ai, Leigh:1997sj}.

\acknowledgments

I thank the anonymous referee of Ref. \cite{Joseph:2013bra} for suggesting the problem. I am grateful to Simon Catterall and Joel Giedt for introducing me to the ideas of exact lattice supersymmetry. I also would like to thank Poul Damgaard, Tom DeGrand and Mithat \"Unsal for a close reading of the manuscript and comments, and especially So Matsuura for some helpful questions and suggestions. This work was supported in part by the Deutsche Forschungsgemeinschaft (SFB/TR 9).

\end{document}